\documentstyle[11pt,paspconf]{article}

\begin{document}

\title{Pushing Back Studies of Galaxies Toward the Dark Ages: High-Redshift
    Ly$\alpha$ Emission-Line Galaxies in the Field}

\author{Esther M. Hu}
\affil{Institute for Astronomy, University of Hawaii, 2680 Woodlawn Drive,
    Honolulu, HI 96822, USA}

\begin{abstract}
Ly$\alpha$ emission presents a means of identifying very high redshift
galaxies, provided such objects can be reliably distinguished from
foreground emission-line galaxies.  Here we report on a program of
imaging and spectroscopic studies being conducted with the Keck
telescope, in concert with additional efforts from the ground and with
HST, to study such high-redshift objects over the range from $z=3\to7$.
The first deep narrow-band imaging studies with Keck are used to show
that strong Ly$\alpha$ emitters, such as have been seen in high-redshift
quasar fields at $z>4.5$, can be found in the field populations. Based
on studies of the HDF and the Hawaii Survey Fields these are
significant contributors to the star formation rates at these epochs.
\end{abstract}

% Keywords should be included, but they are not printed in the hardcopy.


\section{Introduction}

Recent attempts to identify and study galaxies at the high end of the
currently known redshift range ($z\sim 4\to5$) have
primarily focused on color-selection techniques, which rely on the
strong depression of the continuum below the redshifted Lyman break to
roughly select out candidates by redshift.  This method, suggested by
Cowie (1988) and so successfully applied by Steidel and co-workers
(1996) to galaxies up to redshifts of 3.5, has been recently extended
to $z\sim 4.92$ by Franx et al.\ (1997) in the case of gravitationally
lensed objects.

However, the first objects found at $z>4$ were seen to have strong Ly$\alpha$
emission, and could be selected by this property.  In the case of
BR1202--0725$e$, the close companion to a $z=4.69$ quasar, both Ly$\alpha$
emission (Hu et al.\ 1996, Petitjean et al.\ 1996) and color selection (Fontana
et al.\ 1996) were capable of distinguishing the source as a high-redshift
object.  However, other high-$z$ Ly$\alpha$ emitters found in quasar fields,
but well-separated from the quasars, were found with extremely weak continuua,
and could not have been found by color-selection methods (Hu \& McMahon 1996).
This suggests that these populations may also be prevalent in the field samples.

The Lyman alpha searches are complementary to the color techniques for
two reasons. First, at higher redshifts ($z>5$), (unlensed) continuum
magnitudes become very faint, and the identification and spectroscopic
confirmation of color-selected galaxies becomes progressively more
difficult. Here direct selection of Ly$\alpha$ emitters represents our best
chance of finding and mapping a large sample of high-$z$ galaxies.
Secondly, even at more moderate redshifts ($z=2\to4$) where such
objects have been seen (e.g., Pascarelle et al.\ 1996), the
Ly$\alpha$ emitters may represent earlier and less reddened stages in the
galaxy formation process, and measured luminosities, sizes, velocity
widths and clustering as a function of the line's equivalent width may
provide crucial insights into very early galaxy growth.

The present paper reports on an ongoing program to study very high redshift
galaxies selected by their Ly$\alpha$ emission properties, and also demonstrates
the first use of narrow-band filters on the new large (8- $\to$ 10-m class)
telescopes.  Such objects are significant contributors to the integrated star
formation in the field galaxy population by redshifts $z\sim 3.4$ (Cowie \& Hu 
1998).

\section{Narrow-band imaging searches at $z=3\to5$}

As a first step in investigating high-$z$ emitters in the field, we
undertook narrow-band searches of well-studied fields in the HDF and
Hawaii Galaxy Surveys.  The extensive deep, multi-color imaging and
spectroscopic coverage of these fields from the ground and with $HST$
make them ideal testbeds for: (1) comparing methodologies and
sensitivities of emission-line vs continuum color-break techniques for
identifying high-redshift galaxies, (2) determining relative numbers of
foreground emission-line objects and determining the best methods for
distinguishing between (Ly$\alpha$, [O\thinspace{\sc{ii}}], H$\beta$, 
[O\thinspace{\sc{iii}}]) emitters, (3) estimating the relative surface
density of high-redshift candidates as a function of flux from each
technique, and (4) establishing a baseline for comparing blank-field and
targeted searches around high-$z$ objects (such as radio galaxies, DLAs,
and quasars), and for the evolution of galaxy properties at higher
redshifts.  Such studies then permit fine-tuning the search procedures.
Two filters, corresponding to Ly$\alpha$ at $z\sim 3.4$ and 4.6, were
chosen to begin these studies in redshift ranges overlapping color studies
in the literature and by our collaborators (Giallongo et al.\ 1997).

In Fig.~\ref{fig:1} we show a 2-hour exposure on the Hubble Deep Field (HDF)
obtained using the 5390 \AA\ (77 \AA\ wide) filter on LRIS.  The
narrow-band image is compared with an ultra-deep $V$-band image which was
also obtained on LRIS, and covers an area of $380'' \times 275''$, with
FWHM of ${\sim0^{\prime\prime}\kern-2.1mm .\kern+.6mm}7$ and a $1\ \sigma$
limiting narrow-band magnitude of $N(AB)$=26.8 in a $3''$ diameter aperture,
corresponding to a  $1\ \sigma$ flux limit of $6 \times 10^{-18}\ {\rm
erg\ cm}^{-2}\ {\rm s}^{-1}$.

There are 719 objects in the field above the $5\ \sigma$ limit, of which
222 have redshifts in the literature.  Five of the objects in the HDF
narrow-band sample have observed equivalent widths in excess of 100
\AA\, and are circled in Figure \ref{fig:1}. This selection immediately recovers
the one known object lying within the redshift interval, the $z=3.430$
object hd2-06980-1297 (shown in heavy circle), which was identified by
Lowenthal et al.\ (1997) based on emission and absorption
lines in an LRIS spectrum.

\begin{figure}
\vbox to498pt{\rule{0pt}{498pt}}
%\plotfiddle{hue1.eps}{498pt}{0}{73.3}{73.3}{-224}{-30}
\caption{Deep Keck LRIS images of the HDF in $V\/$ ({\it top}) and through a
narrow-band filter centered at 5390 \AA\ ({\it bottom}).  Circles show strong
emitters, which are candidate Ly$\alpha$ galaxies.  This recovers the one
object ({\it heavy circle}, $z=3.430$) with known Ly$\alpha$ in the filter
bandpass.\label{fig:1}}
\end{figure}

\begin{figure}
\plotfiddle{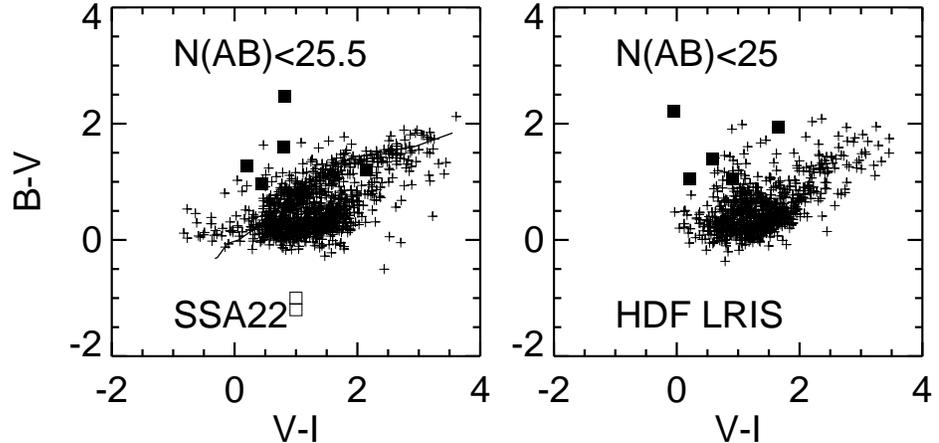}{150pt}{0}{58}{58}{-195}{-60}
\caption{($B-V$\/) vs ($V-I$\/) color-color plots for objects in the HDF and
SSA22 fields selected from the narrow-band catalogues ($N(AB) < 25$ for the
LRIS HDF field and $N(AB) < 25.5$ for the LRIS SSA22 field).  The colors
of emission-line objects in these magnitude-selected catalogues with
equivalent widths in excess of 100 \AA\ in the observed frame are indicated
with squares, and lie at red ($B-V$\/) and blue ($V-I$\/) colors.
Two of the strong emission-line objects in the SSA22 field have
continuua which are either undetected or too faint to provide color
measurements.  They are indicated schematically by open squares
placed at nominal color positions.\label{fig:2}}
\end{figure}

Of the eleven objects with equivalent widths in excess of 100 \AA\ identified 
in this way in the HDF and SSA22, two are completely absent in the continuum 
$B,V,R,I,J,K$ images ($V>27$) and have observed equivalent widths in excess of 
500 \AA, but the remainder have colors which would place them at high redshift.
In Fig.~\ref{fig:2} we plot the locus of colors for the narrow-band selected
objects in the two fields.
The objects that meet our equivalent width criterion ({\it solid squares}) have
predominantly red  ($B-V$\/) colors, but are blue (or close to flat-spectrum
$f_{\nu}$) in ($V-I$) colors, as would be expected for high-redshift
objects that are dominated by star formation (Cowie 1988, Steidel et
al.\ 1996, Lowenthal et al.\ 1997) and which have strong Lyman alpha
forest absorption in the $B$ band.  These objects occupy the same
region of the color-color diagram as objects seen in the tail of the
color-color distribution for the HDF proper, and which correspond to 
spectroscopically identified high-redshift galaxies selected by color.

\begin{figure}
\plotfiddle{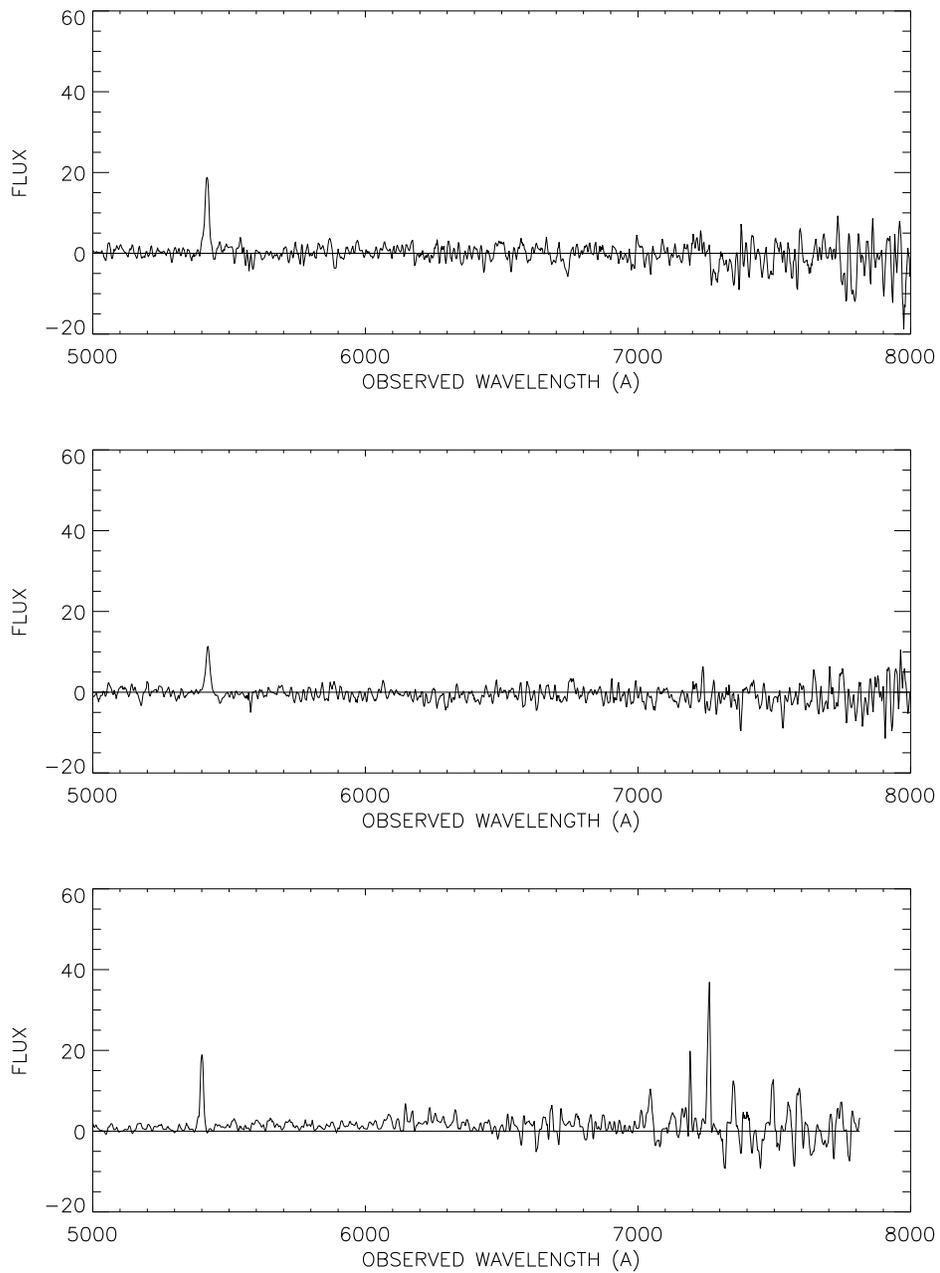}{480pt}{0}{73}{73}{-195}{0}
\caption{Spectra of SSA22 strong Ly$\alpha$ emitters without detected
$V$ continuua ({\it top two panels}) compared with the spectrum of
an [O\thinspace{\sc{ii}}] emitter with extremely high equivalent 
width ({\it bottom panel}) shifted so the [O\thinspace{\sc{ii}}] emission 
lies at the narrow-band wavelength.  The H$\beta$ and [O\thinspace{\sc{iii}}] 
complex is clearly evident in the case of the [O\thinspace{\sc{ii}}] emitter, 
and these cases may be easily distinguished from genuine high-$z$ Ly$\alpha$ 
emission objects.\label{fig:3}}
\end{figure}

\begin{figure}
\plotfiddle{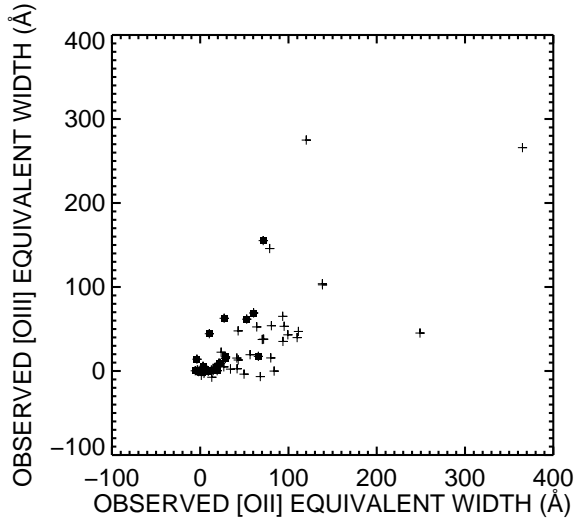}{175pt}{0}{55}{55}{-140}{0}
\caption{Plot of [O\thinspace{\sc{iii}}] vs [O\thinspace{\sc{ii}}] equivalent 
widths ({\it pluses}) from the magnitude-selected Hawaii Survey Fields. 
Overplotted points ({\it solid boxes}) show the $z<0.5$ emitters.\label{fig:4}}
\end{figure}

The objects which are not detected in the continuum lie in the SSA22 field.
Because the equivalent widths in these objects are so high, the lines are 
unlikely a priori to be [O\thinspace{\sc{ii}}] or [O\thinspace{\sc{iii}}], 
where rest equivalent widths generally do not exceed 100 \AA. However, in order 
to check
this further, we obtained spectra of the objects in August of this
year using the LRIS spectrograph in the multi-object mode. The spectra of the
two objects are shown in Fig.~\ref{fig:3} where we compare them with
the highest equivalent width [O\thinspace{\sc{ii}}] line emitter found in the 
Hawaii spectrographic surveys. The [O\thinspace{\sc{ii}}] galaxy, which is 
characteristic of all the spectra with high [O\thinspace{\sc{ii}}] equivalent 
widths (cf., Fig.~\ref{fig:4}), has extremely strong [O\thinspace{\sc{iii}}] 
and H$\beta$, which are features not present in the candidate Ly$\alpha$
emitters. Similarly, the more unlikely possibility that the line is
[O\thinspace{\sc{iii}}] is ruled out by the absence of the H$\alpha$ line and 
the [O\thinspace{\sc{iii}}]
and H$\beta$ complex. There therefore appears little doubt that these
objects are indeed Ly$\alpha$ emitters. We may also note the
the absence of strong {{\rm C}\thinspace{\sc{iv}}} in the spectra.

\begin{figure}[t] 
\plotfiddle{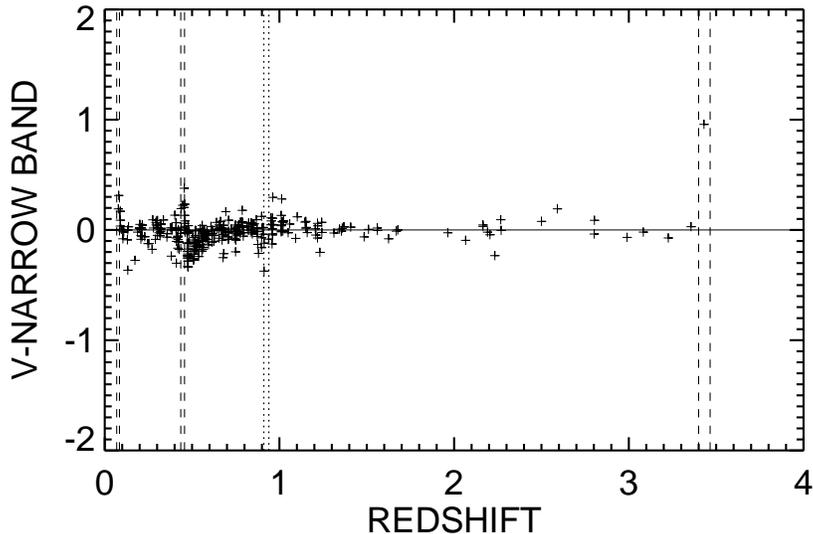}{185pt}{0}{55}{55}{-210}{-30}
\caption{Excess emission in narrow-band over $V$-band vs redshift
from objects in the HDF and SSA22 fields.  Objects with spectroscopic
identifications in the literature over the HDF and SSA22 fields sampled
by the LRIS deep imaging fields are shown here.  The positions and redshift 
ranges corresponding to features such as Mg\thinspace{\sc{ii}} ({\it dotted 
lines}) and [O\thinspace{\sc{iii}}], [O\thinspace{\sc{ii}}], and Ly$\alpha$ 
({\it dashed lines}) falling within the filter bandpass are indicated.  The 
structure of the continuum near the [O\thinspace{\sc{ii}}] feature may be seen 
reflected in both the `dip' and dispersion of points in this region, and it 
may also be seen that emission-line objects in [O\thinspace{\sc{ii}}] and a 
few [O\thinspace{\sc{iii}}] emitters may be detected.  Underlying continuum 
features can also increase the dispersion of points in different redshift 
regions.\label{fig:5}}
\end{figure}

The extensive redshift information which is available for the fields also
allows us to check the equivalent width selection procedure.  In
Fig.~\ref{fig:5} we show ($V-N$\/) vs.\ redshift, with the redshift intervals
matching the narrow-band filter band-pass indicated for emission from
[O\thinspace{\sc{iii}}] $\lambda\,5007$ (at $z\sim 0.08$),
[O\thinspace{\sc{ii}}] $\lambda\,3727$ (at $z\sim 0.45$),
Mg\thinspace{\sc{ii}} $\lambda\,2800$ (at $z\sim 0.93$), and Ly$\alpha$
$\lambda\,1216$ (at $z\sim 3.43$).  These are the most prominant spectral
features in most galaxies, with Mg\thinspace{\sc{ii}} normally seen in
absorption, and they are clearly seen in the color-redshift plot.  The
scatter of points gives an idea of the real distribution of the signal
strength of each feature, and it may also be seen that the high-$z$
Ly$\alpha$ emitter, corresponding to the Lowenthal et al.\ object, does
indeed have appreciably higher equivalent width than any of the
[O\thinspace{\sc{ii}}] emitters, though it is in fact one of the lowest
equivalent width objects in the sample.  Other structures arise from
additional spectral features ({\it e.g.}, the 4000 \AA\ break and
[Fe\thinspace{\sc{ii}}] absorption lines), which add to the dispersion.

As we move to higher redshift ($z=4.6$) we continue to see high equivalent
emitters ($W>100$ \AA\ in the observed frame).  A 2-hr exposure taken through
the 6740/80 \AA\ narrow-band filter produced 3 such objects with $AB <
24.5$.  The lowest equivalent width object is an [O\thinspace{\sc{ii}}]
emitter which has a strong [O\thinspace{\sc{iii}}] and H$\beta$ complex,
while the two higher equivalent width objects are Ly$\alpha$ emitters.  In
the spectroscopic studies, probable [O\thinspace{\sc{ii}}] emitters are also
targeted for followup spectra, in order to understand the differences between
emission-line selection and continuum selection.  This will be particularly
important to make a convincing case for objects where the corresponding
[O\thinspace{\sc{ii}}] line cannot be detected (e.g., at $z\ga5.2$), in
contrast to the case for the $z=4.5\to4.7$ Ly$\alpha$ emitters (cf., Hu et
al.\ 1997; Egami 1998).

In the three fields that we have observed so far there are approximately
six Ly$\alpha$ emitters per 23 $\sq'$ field or about $1,000/\sq^{\circ}$
lying in the redshift range $z=3.405 \to 3.470$, or roughly
15,000/unit $z/\sq^{\circ}$.  Most of the objects detected
in the continuum have $V < 25.25$.
To this $V=25.25$ magnitude limit a color criterion ($B-V$\/) $> 1$,
$(V-I) < 1.6$ of the type discussed above gives 80 objects  per
sample area, or assuming a $z$ range of $3.1 \to 3.5$, where the
upper limits is determined by the passage of the forest through the
red edge of the $V$ band, a surface density of 15,000/unit $z/\sq^{\circ}$.
While the number is quite uncertain because of the small number statistics,
the choice of magnitude cutoff and the possibility of redshift clustering, a
large fraction -- very roughly half -- of the $B$ drop-out galaxies are
strong Ly$\alpha$ emitters.  This is roughly consistent with the Lyman alpha
emitting fraction of the color selected galaxies with measured high redshifts,
again giving us confidence that the methodology is sound.

The rest frame equivalent widths of the systems are consistent with
stellar excitation for an initial mass function dominated by massive stars
(Charlot \& Fall 1993).  If the observed emission were primarily due to
stars, and there were no internal scattering and extinction, then a
luminosity of $10^{42}$ erg s$^{-1}$ would correspond to a star formation
rate in solar masses ($M_{\odot}$) per year of $\sim1$ $M_{\odot}$
yr$^{-1}$, where we use Kennicutt's (1983) relation between H$\alpha$
luminosity and star formation rate (SFR) of SFR = $L($H$\alpha$
$\times\ 8.9\times 10^{-42}$ erg s$^{-1}$ $M_{\odot}$ yr$^{-1}$, and
assume a ratio of Ly$\alpha$ to H$\alpha$ (8.7) that applies for Case B
recombination.  Given the Hubble time at these redshifts of $7\times
10^8\ h^{-1}$ yr ($q_0=0.5$ for $z=4.6$) the integrated amount of star
formation in the objects is clearly small compared to that of a `normal'
$L^*$ galaxy with $6\times 10^{10}\ h^{-1}$ $M_{\odot}$ of stars.

\begin{figure}[t]
\plotfiddle{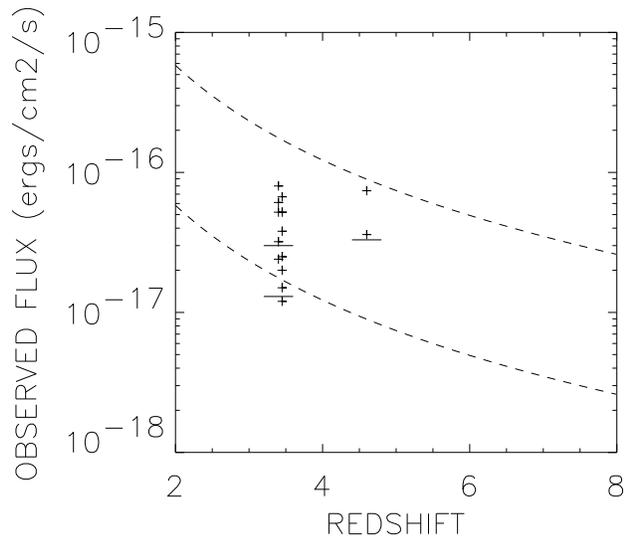}{185pt}{0}{55}{55}{-130}{0}
\vskip0.2in
\caption{Crosses show the line fluxes of the objects at $z=3.4$
in three fields and at $z=4.6$ in the SSA22 field only.
The solid lines show the 5 sigma detection limits for bare line
emitters for the 2-hour exposure at 5390 \AA\ in the HDF and the
5-hour exposure in the SSA22 field and the 2-hour exposure
at 6740 \AA\ in the SSA22 field. (Some objects lie slightly
below these limits because of the continuum subtraction.)
There is relatively little
wavelength dependence in the sensitivity limit at the same exposure
time because the filters have been positioned
in relatively line free regions of the night sky. The dashed lines
show the expected fluxes for objects with $1\times 10^{42}$ erg~s$^{-1}$
and $1\times 10^{43}$ erg~s$^{-1}$
for $H_0=65$ km~s$^{-1}$ Mpc$^{-1}$ and $q_0=0.5$.
For dust-free objects with
Salpeter mass functions these Ly$\alpha$ luminosities correspond
to 1 and 10 $M_{\odot}$ per year respectively.\label{fig:6}
}
\end{figure}

\section{Next steps}

The next stage in the project 
is to develop much larger samples at the two redshifts already covered
by the protoyping project and to extend the work into the $z>5$
range using longer wavelength filters. Figure~\ref{fig:6} provides an
overview of what we require in both these areas by showing
the line fluxes of the existing objects at the two redshifts
together with the sensitivity limits and the expected decrease in
the fluxes as a function of redshift for $q_0=0.5$ models. These
data wll be combined with ongoing deep multi-color optical/IR studies for both
field samples and high-$z$ quasar fields (in upcoming HST observations),

Searches for yet higher $z$ galaxies are of necessity forced into the near 
IR since
the light below $1216(1+z)~{\rm\AA}$\ will be essentially extinguished by the
strong Ly~$\alpha$\ forest blanketing or by H~I Gunn-Peterson at the redshifts
where the IGM was neutral.  Objects at these high redshifts will also be
extremely faint:  even for $q_0 = 0.5$\ a galaxy burning $10^{10}~{\rm
M}_{\odot}$\ of baryons in a Hubble time with a Salpeter IMF will have an
$AB$\ magnitude of roughly 26.5 (100 nJy) --- almost irrespective of
redshift --- at wavelengths corresponding to the rest ultraviolet
($1216~{\rm\AA} - 300~{\rm\AA}\,(1+z)$) (e.g., Cowie 1988).  For $q_0 = 0$\ the
result is redshift dependent, but such a galaxy at $z = 8$\ would be almost an
order of magnitude fainter.  Impulsive burning could increase the flux a
little but probably not by more than an order of magnitude even in the open
case.  Since characteristic baryon masses in these early objects are likely to
be about $10^8 - 10^9 {\rm M}_{\odot}$\ 
we will probably have to reach nJy level  fluxes to detect
populations at these redshifts.  We can look forward to detecting these
objects with NGST.

\acknowledgments
This work was supported in part by NASA grants GO-6222.01-95A and
AR-6377.06-94A from Space Telescope Science Institute.%, which is 
%operated by AURA, Inc. under NASA contract.

\end{document}